\documentclass[twocolumn,amsmath,amssymb,aip,pop]{revtex4}

\usepackage{graphicx}
\usepackage{bm}
\usepackage{amsfonts}

\usepackage[utf8]{inputenc}

\begin{document}

\title{Self-consistent equilibrium of a helical magnetic flux rope in a finite-pressure plasma}

\author{Oleg K. Cheremnykh}
\affiliation{Space Research Institute, Pr. Glushkova 40 k.4/1,
Kyiv 03187, Ukraine}

\author{Viktor Fedun}
\affiliation{Plasma Dynamics Group, Department of Automatic
Control and Systems Engineering, The University of Sheffield,
Sheffield, S1 3JD, United Kingdom}

\author{Volodymyr M. Lashkin}
\email{vlashkin62@gmail.com} \affiliation{$^1$Institute for
Nuclear Research, Pr. Nauki 47, Kyiv 03028, Ukraine}
\affiliation{$^2$Space Research Institute, Pr. Glushkova 40 k.4/1,
Kyiv 03187,  Ukraine}


\begin{abstract}
We present an analytical model of  the self-consistent equilibrium
of a magnetic flux rope which is obtained in cylindrical geometry.
The  equilibrium azimuthal magnetic field and plasma pressure are
determined in a  self-consistent way through the current density
which is derived as a solution of a nonlinear equation.  By
minimizing the energy functional, it was shown that the
constrained equilibrium state is stable. The obtained results are
also applicable to the cylindrical tokamak magnetic
configurations. It is shown that the analytically predicted radial
profiles of equilibrium quantities are in good agreement with the
experimental data.
\end{abstract}

\maketitle

The concept of a helical magnetic flux rope is widely used in
solar physics \cite{Rytova2015-book-1}. By some estimates
\cite{Stenflo2017-2} about 90~\% of the magnetic flux on the
surface of the Sun can be concentrated in magnetic flux ropes.
These structures play a dominant role in the  flare energy release
and in the formation of coronal mass ejections
\cite{Uchida1996-3,Priest2000-book-4,Aschwanden2019-book-5,Daughton2011POP,Nakamura2016POP,Chen2017POP}.
Usually the magnetic flux rope is considered as a slightly curved
plasma cylinder of radius $a$ in the cylindrical coordinate system
$r,\varphi,z$ with a sufficiently large radius of curvature, along
which electric current flows. Under such assumptions, one can
neglect the toroidal corrections in the distribution of
equilibrium quantities inside the plasma column and consider it
cylindrically symmetrical. It is also assumed that inside and
outside the plasma column there is a constant magnetic field
parallel to the current. For a long time when considering the
stability of plasma ropes the question arises about the presence
of an azimuthal magnetic field at distances greater than $a$. Note
that, generally speaking, there are models of a magnetic rope in
which there is no azimuthal magnetic field outside the radius of
the rope cross section \cite{Zhelyazkov2015-6,Solovev2014-7}, as
well as models in which this field is present
\cite{Liu2017-10,Titov2018-11,Cheremnykh2018-12,Zaitsev2020-13}.
In this case, the corresponding models of the equilibrium of
magnetic rope are considered. The question of the behavior of the
azimuthal magnetic field beyond the radius of the magnetic rope
was discussed in detail by Parker \cite{Parker2007-book-14} and
Melrose \cite{Melrose2017-15}. In a recent paper
\cite{Solovev2021-16}, in the cylindrical equilibrium model, it
was stated that the azimuthal magnetic field  $B_{\varphi}$
vanishes at the rope boundary $r=a$ and is absent for $r>a$. In
fact, this condition is the definition of the radius of the
magnetic rope. Wherein  it was shown that the total current
flowing through the rope is zero. For such a postulated behavior
of the azimuthal magnetic field, the authors of
Ref.~\cite{Solovev2021-16} found equilibrium states with zero
pressure and total current inside the rope. On the other hand, the
question of the physical realization of such states remains open,
since taking into account the plasma pressure, which corresponds
to the real situation, can lead to the destruction of the
constructed equilibrium states. This example, in particular,
indicates the need to develop approaches that allow one to find a
self-consistent equilibrium that would not depend on any
assumptions about the behavior of equilibrium quantities. In this
Communication, we propose a method by which it was possible to
analytically find a self-consistent cylindrically symmetric
equilibrium in a helical magnetic filament. The integral equation
of the pinch effect is added to the two commonly used local
equilibrium equations \cite{Kadomtsev1988-17}. Another local
equation follows from the latter, and as a result we obtain a
nonlinear equation which makes it possible to uniquely determine
the equilibrium state. All equilibrium quantities turn out to be
distributed along the radius to infinity. However, a significant
part of the current is concentrated in a small neighborhood of the
$z$ axis. We show that the found equilibrium corresponds to the
state of the rope with the minimum energy and, therefore, is
stable. The proposed model of the magnetic flux rope can easily be
generalized to the case of a cylindrical tokamak
\cite{Kadomtsev1988-17,Miyamoto-18}. The latter has been studied
in detail experimentally and theoretically. We show that the
results obtained, in particular, the radial dependence of the
plasma pressure, are in good agreement with the tested
experimental data of different types of tokamaks.

We consider an inhomogeneous equilibrium state of an ideal plasma,
which is established self-consistently when a direct current flows
through it, and assume that in a cylindrical coordinate system
$r,\varphi,z$ this state has cylindrical symmetry about the $z$
axis, so that all equilibrium quantities depend only on the radial
coordinate $r$. We assume that the plasma is immersed in a uniform
constant magnetic field $B_{z}$ directed along the $z$ axis, and
an inhomogeneous axisymmetric current $I$ flows in the $z$
direction with a maximum current density $j_{z}^{0}$ at $r=0$. The
current density $j_{z}(r)$ is related to the azimuthal magnetic
field $B_{\varphi}$ by the Ampere equation for the case of axial
symmetry,
\begin{equation}
\label{main-1}
j_{z}=\frac{c}{4\pi}\frac{1}{r}\frac{d}{dr}(rB_{\varphi}).
\end{equation}
The plasma equilibrium condition has the form
\begin{equation}
\label{equilibrium-2} \frac{dp}{dr}+\frac{1}{c}j_{z}B_{\varphi}=0,
\end{equation}
where $p$ is the plasma pressure, $c$ is the speed of light. If
$j_{z}=0$, Equations (\ref{main-1}) and (\ref{equilibrium-2})
describe the simplest equilibrium state, which is usually used
when considering linear MHD waves: plasma with uniform pressure
and magnetic field fills the entire space uniformly
\cite{Kadomtsev1988-17}. In the presence of current, Eqs.
(\ref{main-1}) and (\ref{equilibrium-2}) contain three
interrelated quantities $p$, $j_{z}$ and $B_{\varphi}$  with
arbitrary, generally speaking, radial profiles. An unambiguous
solution of these equations can be obtained by adding one more
equation for the equilibrium quantities. We will find this
equation from the well known pinching condition (integral
equilibrium condition). We represent the plasma pressure in the
form $p=\tilde{p}+p_{\infty}$, where $\tilde{p}$ is responsible
for the inhomogeneous part, and $p_{\infty}$  corresponds to the
homogeneous part. The geometry of the problem is sketched in
Fig.~1. From Eqs. (\ref{main-1}) and (\ref{equilibrium-2}) it
follows that
\begin{equation}
\frac{d\tilde{p}}{dr}r^{2}=-\frac{1}{8\pi}\frac{d}{dr}(rB_{\varphi})^{2}.
\end{equation}
Integrating this equation over $r$ from $0$ to $\infty$, we have
\begin{equation}
\label{integral} \int_{0}^{\infty}\frac{d\tilde{p}}{dr}r^{2}dr=
\left.-\frac{1}{8\pi}(rB_{\varphi})^{2}\right|_{0}^{\infty}.
\end{equation}
According to Eq. (\ref{main-1}), in the limit $r\rightarrow\infty$
we have $rB_{\varphi}=2I/c$, where
$I=2\pi\int_{0}^{\infty}j_{z}rdr$ is the total current flowing
through the plasma, and then Equation (\ref{integral}) becomes
\begin{equation}
\label{before-parts}
\int_{0}^{\infty}\frac{d\tilde{p}}{dr}r^{2}dr= -\frac{I^{2}}{2\pi
c^{2}}.
\end{equation}
Integrating left hand side of Eq.~(\ref{before-parts}) by parts,
one can obtain
\begin{equation}
\label{parts}
\left.\tilde{p}r^{2}\right|_{0}^{\infty}-2\int_{0}^{\infty}\tilde{p}rdr=-\frac{I^{2}}{2\pi
c^{2}}.
\end{equation}
Assuming that $\tilde{p}$ is bounded at $r=0$ and vanishes at
infinity faster than $1/r^{2}$, we get from Eq. (\ref{parts}) the
pinch effect equation,
\begin{equation}
\label{pinch1} 2\int_{0}^{\infty}\tilde{p}rdr=\frac{I^{2}}{2\pi
c^{2}},
\end{equation}
which can be written in the form
\begin{equation}
\label{pinch2}
2\pi\int_{0}^{\infty}\left(\tilde{p}-\frac{j_{z}I}{2c^{2}}\right)rdr=0.
\end{equation}
\begin{figure}
\includegraphics[width=2.5in]{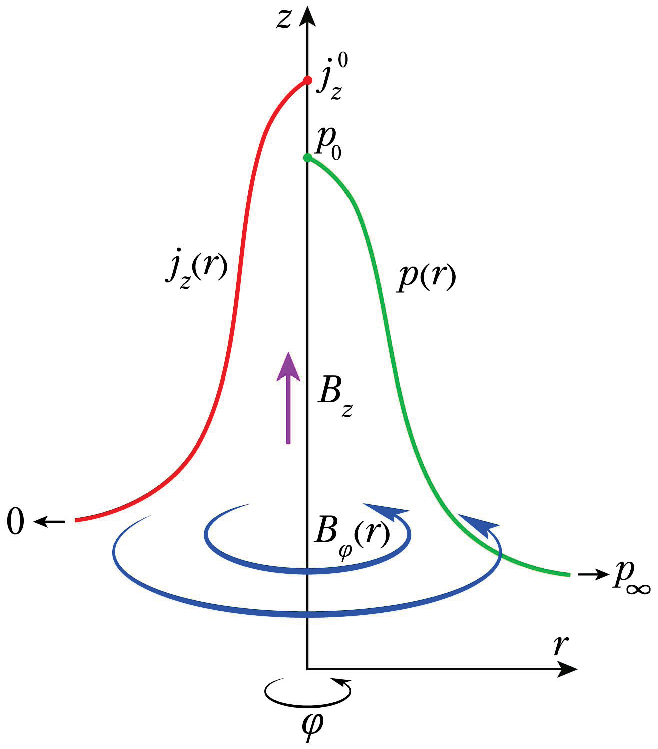}
\caption{\label{fig1} Sketch of the geometry of the problem in
cylindrical coordinates $(r,\varphi,z)$.}
\end{figure}
This equation will be satisfied if the plasma pressure $p$ and
current density $j_{z}$ satisfy the condition
\begin{equation}
\label{condition2-5} p=\alpha j_{z}+p_{\infty},
\end{equation}
where $\alpha=I/(2c^{2})$. Equation (\ref{condition2-5})  together
with Eqs.~(\ref{main-1}) and (\ref{equilibrium-2}) makes it
possible to self-consistently find all equilibrium quantities,
that is, the azimuthal magnetic field, plasma pressure and current
density, and determine the corresponding radial profiles. From
Eqs. (\ref{main-1}), (\ref{equilibrium-2}) and
(\ref{condition2-5}) one can obtain a nonlinear equation for the
current density,
\begin{equation}
\label{nonlin1-7}
\frac{1}{r}\frac{d}{dr}\left(\frac{r}{j_{z}}\frac{dj_{z}}{dr}\right)+\frac{4\pi}{\alpha
c^{2}}j_{z}=0,
\end{equation}
The solution of this equation with boundary conditions
$j_{z}=j_{z}^{0}$ at $r=0$ and $j_{z}\rightarrow 0$ as
$r\rightarrow\infty$ gives the radial dependence of the current
density and makes it possible to completely solve the problem of
self-consistent equilibrium. As will be shown below, this solution
can be extended to the case when the plasma is mainly contained in
a cylinder with a certain radius $a$. Such a cylinder can be
considered as a helical magnetic rope. Representing the current
density in the form
\begin{equation}
\label{current-density-9} j_{z}=j_{z}^{0}\exp \psi,
\end{equation}
where $\psi$ is an arbitrary function of radius $r$ that satisfies
the boundary conditions $\psi=0$ at $r=0$ and $\psi\rightarrow
-\infty$ as $r\rightarrow\infty$, we obtain from Eq.
(\ref{nonlin1-7}),
\begin{equation}
\label{nonlin2-11}
\frac{1}{r}\frac{d}{dr}\left(r\frac{d\psi}{dr}\right)+\beta\exp
\psi=0,
\end{equation}
where $\beta=4\pi j_{z}^{0}/(\alpha c^{2})$. Taking into account
these boundary conditions, we look for the function $\psi$ in the
form
\begin{equation}
\label{solution-try-15} \psi=\ln \frac{1}{(1+\gamma
r^{2})^{\delta}},
\end{equation}
where $\gamma$ and $\delta$ are still unknown constants.
Substituting Eq.~(\ref{solution-try-15}) into
Eq.~(\ref{nonlin2-11}) one can readily get $\gamma=\beta/8$ and
$\delta=2$, so that we finally find
\begin{equation}
\label{solution-main-20} \psi=\ln
\frac{1}{(1+r^{2}/r^{2}_{\ast})^{2}},
\end{equation}
where we have introduced the effective radius of the magnetic rope
as $r_{\ast}=2\sqrt{2}/\beta$. The radial dependence of the
current density is found from Eqs.~(\ref{current-density-9}) and
(\ref{solution-main-20}),
\begin{equation}
\label{current-density-21}
j_{z}=\frac{j_{z}^{0}}{(1+r^{2}/r^{2}_{\ast})^{2}}.
\end{equation}
In turn, from Eqs.~(\ref{main-1}) and (\ref{condition2-5}) for the
azimuthal magnetic field one can write
\begin{equation}
B_{\varphi}=-(\alpha c/j_{z})dj_{z}/dr,
\end{equation}
and then from Eq.~(\ref{current-density-21}) we have
\begin{equation}
\label{azimuthal-field-23}
 B_{\varphi}=\frac{B_{\varphi}^{0}r/r_{\ast}}{(1+r^{2}/r^{2}_{\ast})},
\end{equation}
where we have introduced the notation $B_{\varphi}^{0}=2\pi
r_{\ast}j_{z}^{0}/c$. Note that the radial dependence of
$B_{\varphi}$ is not monotonic. The azimuthal magnetic field
increases from zero at $r=0$, reaches a maximum value
$B_{m}=B_{\varphi}^{0}/2$ at $r_{m}=r_{\ast}$ and then decreases
at infinity to zero as $\sim 1/r$. The plasma pressure $p$ can be
found from Eq.~(\ref{condition2-5}),
\begin{equation}
\label{pressure-25}
p=\frac{p_{0}}{(1+r^{2}/r^{2}_{\ast})^{2}}+p_{\infty},
\end{equation}
where $p_{0}=(B_{\varphi}^{0})^{2}/8\pi$. Radial profiles of the
azimuthal magnetic field $B_{\varphi}$, plasma pressure $p$, and
current density $j_{z}$ are shown in Fig.~2.
\begin{figure}
\includegraphics[width=3.3in]{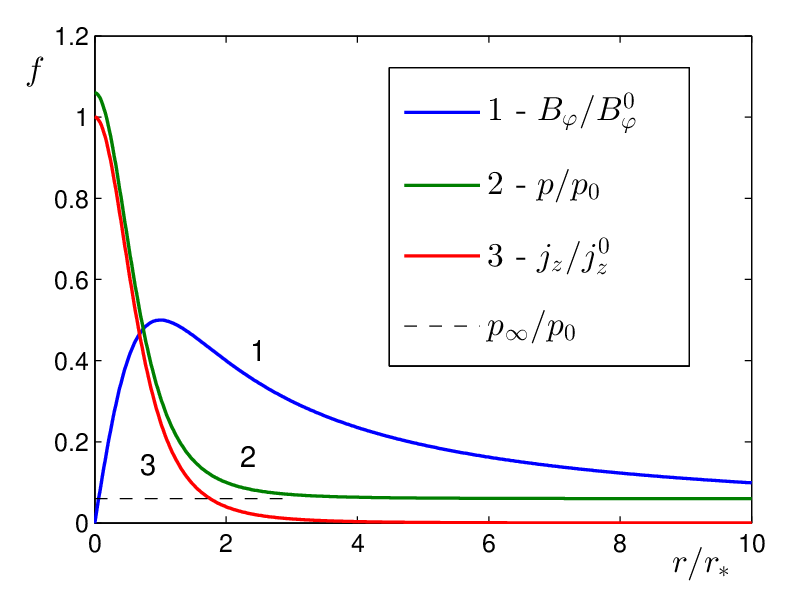}
\caption{\label{fig2} Radial dependences of the normalized
equilibrium quantities $f$: azimuthal magnetic field
$B_{\varphi}$, plasma pressure $p$, and current density $j_{z}$.}
\end{figure}
From Eqs.~(\ref{main-1}) and (\ref{azimuthal-field-23}) we obtain
the total current $I$ flowing in the plasma,
\begin{equation}
\label{full-current-27} I=\int_{0}^{\infty}j_{z}2\pi r
dr=\frac{c}{4\pi}\int_{0}^{\infty}\frac{1}{r}\frac{d}{dr}(rB_{\varphi})2\pi
r dr=\pi j_{z}^{0}r_{\ast}^{2}.
\end{equation}
Then $r_{\ast}$ can be considered as the radius of the cylinder,
inside which  the  current with the uniform density $j_{z}^{0}$
flows. Then the equilibrium quantities (\ref{current-density-21}),
(\ref{azimuthal-field-23}) and (\ref{pressure-25}) can be written
as
\begin{equation}
\label{jz-profile} j_{z}=\frac{j_{z}^{0}}{(1+\pi
j_{z}^{0}r^{2}/I)^{2}},
\end{equation}
\begin{equation}
\label{B-profile} B_{\varphi}=\frac{2\pi j_{z}^{0}r}{c(1+\pi
j_{z}^{0}r^{2}/I)},
\end{equation}
\begin{equation}
\label{p-profile} p=\frac{j_{z}^{0}I}{2c^{2}(1+\pi
j_{z}^{0}r^{2}/I)^{2}}+p_{\infty}.
\end{equation}
All these quantities depend only on $j_{z}^{0}$ and $I$ and are
consistent with each other. The plasma pressure and azimuthal
magnetic field satisfy the equilibrium pressure balance equation
\begin{equation}
\frac{d}{dr}\left(p+\frac{B_{\varphi}^{2}}{8\pi}\right)+\frac{B_{\varphi}^{2}}{4\pi
r}=0,
\end{equation}
which follows from Eqs.~(\ref{main-1}) and (\ref{equilibrium-2}),
and hence are also consistent with each other.

The radial profiles of equilibrium quantities never vanish and we
have to determine what to call the radius of the cross section of
the magnetic flux rope. Intuitively, it can be defined as the
distance at which it is necessary to take into account the current
flowing in the rope. From Eq.~(\ref{jz-profile}) it can be found
that the current flowing inside a cylinder of radius $r$ is
\begin{equation}
I_{z}(r)=I\frac{(r/r_{\ast})^{2}}{1+r^{2}/r_{\ast}^{2}}.
\end{equation}
It can be seen that in order for a significant part of the current
to flow inside this cylinder, it is necessary $r\gg r_{\ast}$ and
then $I_{z}(r)\sim I(1-r_{\ast}^{2}/r^{2})$. For example, in order
for 90 ~\% of the current to flow inside the cylinder, we can put
$r=3r_{\ast}$. Then assuming that the magnetic rope is bounded by
a cylindrical surface of radius $a$ and 90 ~\% of the current
flows inside this surface, we obtain that the radius $a$ is
\begin{equation}
\label{a} a=3r_{\ast}=3\sqrt{I/(\pi j_{z}^{0})}.
\end{equation}
Then, with a given accuracy, we can assume that, at distances
greater than $a$, the azimuthal magnetic field decreases as $\sim
1/r$, the plasma pressure coincides with the currentless plasma
pressure, and there is no electric current.

The physical relevance of  the found equilibrium state depends on
whether it is stable. To study the stability we write the energy
functional as
\begin{equation}
\label{energy1}
E=2\pi\int_{0}^{a}\left(\frac{p-p_{\infty}}{\gamma-1}+\frac{B_{\varphi}^{2}}{8\pi}\right)rdr,
\end{equation}
where $\gamma$ is the ratio of specific heats, the first and
second terms in brackets correspond to the kinetic and magnetic
pressures respectively, and look for a minimum of the energy $E$
subject to the fixed total current
$I=2\pi\int_{0}^{a}j_{z}rdr=\mathrm{const}$. Then
Eq.~(\ref{energy1}) can be written as
\begin{equation}
\label{energy2} E=W+\lambda I+F,
\end{equation}
where $\lambda >0$, and
\begin{equation}
W=\frac{1}{4}\int_{0}^{a}B_{\varphi}^{2}rdr, \quad
F=2\pi\int_{0}^{a}\left(\frac{p-p_{\infty}}{\gamma-1}-\lambda
j_{z}\right)rdr .
\end{equation}
Since $W>0$ and $I>0$, it  follows from Eqs.~(\ref{condition2-5}),
(\ref{jz-profile}) and (\ref{p-profile}) that the minimum of
functional (\ref{energy2}) corresponds to $F=0$, that is, to
self-consistent equilibrium values (\ref{current-density-21}),
(\ref{azimuthal-field-23}) and (\ref{pressure-25}) provided that
$\lambda=\alpha/(\gamma-1)$. In this case, the energy of the
stable equilibrium state is
\begin{eqnarray}
E_{min}=W+\frac{\alpha}{\gamma-1}I
\nonumber \\
=\frac{I^{2}}{c^{2}r_{\ast}^{2}}\int_{0}^{a}
\left(\frac{1}{\gamma-1}+\frac{r^{2}}{r_{\ast}^{2}}\right)\frac{rdr}{(1+r^{2}/r_{\ast}^{2})^{2}}.
\label{energymin}
\end{eqnarray}

Verification of the obtained results is conveniently carried out
according to the tested results obtained in the framework of
similar models. Let us introduce the length of the magnetic rope
as $L=2\pi R$, where $R$ is the major radius, and then the
considered equilibrium will not differ from the equilibrium of the
plasma column in a cylindrical tokamak
\cite{Kadomtsev1988-17,Miyamoto-18}. We use the dimensionless
quantity $q=rB_{z}/(RB_{\varphi})$ \cite{Miyamoto-18}. From
Eq.~(\ref{B-profile}) we get
\begin{equation}
\label{q} q=q_{0}(1+r^{2}/r_{\ast}^{2}),
\end{equation}
where $q_{0}=cB_{z}/(2\pi Rj_{z}^{0})$.  Further, we consider
$q_{0}$ to be a free parameter through which all equilibrium
quantities can be expressed. From Eqs.~(\ref{azimuthal-field-23})
and (\ref{q}) we find
\begin{equation}
\label{beforeSI} \hat{j}_{z}^{0}=\frac{c}{2}\frac{B_{z}}{\pi
q_{0}R},\quad \hat{r}_{\ast}^{2}=\frac{2}{c}\frac{q_{0}IR}{B_{z}},
\quad
\hat{B}_{\varphi}^{0}=\sqrt{\frac{2}{c}\frac{IB_{z}}{q_{0}R}},
\end{equation}
which in SI units can be rewritten as
\begin{equation}
\label{SI} \hat{j}_{z}^{0}=\frac{5B_{z}}{\pi q_{0}R},\quad
\hat{r}_{\ast}^{2}=\frac{q_{0}IR}{5B_{z}}, \quad
\hat{B}_{\varphi}^{0}=\sqrt{\frac{IB_{z}}{5q_{0}R}}.
\end{equation}
Here,  $B_{z}$ is in tesla (T), $R$ in meter (m), and $I$ in
megaampere (MA). Inserting these expressions into
Eqs.~(\ref{jz-profile}), (\ref{B-profile}) and (\ref{p-profile})
yields
\begin{eqnarray}
q=q_{0}(1+5\rho^{2}), \label{q-43}
\\
j_{z}=\hat{j}_{z}^{0}/(1+5\rho^{2})^{2} , \label{jz-44}
\\
B_{\varphi}=\sqrt{5}\hat{B}_{\varphi}^{0}\rho/(1+5\rho^{2}) ,
\label{B-45}
\\
p=(\hat{B}_{\varphi}^{0})^{2}/8\pi(1+5\rho^{2})^{2}+p_{\infty},
\label{p-46}
\end{eqnarray}
where $\rho=r/\sqrt{q_{0}IR/B_{z}}$. Let us compare
Eqs.~(\ref{q-43})-(\ref{p-46}) with the corresponding equations
for the equilibrium quantities obtained in
Ref.~\cite{Kadomtsev1987-19} within the framework of the
cylindrical tokamak model. In Ref.~\cite{Kadomtsev1987-19}, the
experimental results were taken into account, as well as some
features of plasma equilibrium in tokamaks. For example, it was
taken into account that sawtooth oscillations in the center of the
plasma column give $q_{0}=1$, and also that the plasma column is
surrounded by vacuum, therefore $p_{\infty}=0$. The expressions
obtained in Ref.~\cite{Kadomtsev1987-19} for the current density
and $q$ completely coincide with Eqs.~(\ref{q-43}) and
(\ref{jz-44}) at $q_{0}=1$ . The pressure profile from
Ref.~\cite{Kadomtsev1987-19} also coincides with  Eq.~(\ref{p-46})
at $q_{0}=1$ and $p_{\infty}=0$. From Eq.~(\ref{p-46}) we also
have
\begin{equation}
\label{beta} \beta=\frac{8\pi
p}{B_{z}^{2}}=\frac{\beta_{0}}{(1+5\rho^{2})^{2}},
\end{equation}
where $\beta_{0}=gI/(aB_{z})$ and $g=a/(5q_{0}R)$. The parameter
$\beta_{0}$ is close to the Troyon critical parameter $\beta_{cr}$
\cite{Troyon1999-20} obtained for tokamaks from experimental data,
\begin{equation}
\label{troyon} \beta_{cr}=g_{N}\frac{I}{aB_{z}}, \quad
g_{N}=3.5\times 10^{-2}.
\end{equation}
From Eq.~(\ref{beta}), in particular, it follows that the maximum
equilibrium pressure increases with increasing $a/R$. From
Eqs.~(\ref{energymin}) and (\ref{beforeSI}) we find an expression
for the minimum energy of the plasma column,
\begin{equation}
\label{energymin1}
E_{min}=\frac{B_{z}^{2}}{4R^{2}}\int_{0}^{a}\frac{1}{q^{2}}\left(r^{2}
+\frac{r_{\ast}^{2}}{\gamma-1}\right)rdr.
\end{equation}
This expression, up to the replacement
$r_{\ast}^{2}/(\gamma-1)\rightarrow r_{\ast}^{2}$, coincides with
the corresponding expression of Ref.~\cite{Kadomtsev1987-19},
obtained with the aid of variational analysis. In
Ref.~\cite{Kadomtsev1987-19}, a graphical dependence of the plasma
pressure for a number of tokamaks obtained directly from the
experiment has also been presented. From Fig.~2 it can be seen
that the behavior of the pressure profile in the plasma cylinder
is in good agreement with experimental measurements in plasma
tori. Small deviations are observed at the plasma periphery, where
atomic processes play a dominant role in tokamaks.

In conclusion, we have presented a new model of self-consistent
equilibrium of a magnetic flux rope in cylindrical coordinates. A
nonlinear equation for the current density is obtained and its
analytical solution is found. The remaining equilibrium
quantities, namely, the azimuthal magnetic field and plasma
pressure are determined in a self-consistent way through the
current density. By minimizing the energy functional, we have
shown that obtained equilibrium state is stable. It has been also
shown that the analytically predicted radial profiles of
equilibrium quantities are in good agreement with the experimental
data for a number of tokamaks.

VF and OC would like to thank The Royal Society International
Exchanges Scheme, collaboration with Ukraine  (IES/R1/21117). VF
is grateful to Science and Technology Facilities Council (STFC)
grant ST/V000977/1 for the financial support provided.


\begin{thebibliography}{59}

\bibitem{Rytova2015-book-1}
M.~Ryutova, \emph{Physics of Magnetic Flux Tubes}
(Springer-Verlag, Berlin, 2015).

\bibitem{Stenflo2017-2}
J.~O. Stenflo,  Space Sci. Rev. \textbf{210}, 5 (2017).

\bibitem{Uchida1996-3}
\emph{Magnetodynamic Phenomena in the Solar Atmosphere}, edited by
Y. Uchida, T. Kosugi, and H.~S. Hudson (Kluwer Academic
Publishers, London, 1996).

\bibitem{Priest2000-book-4}
E.~R. Priest, \emph{Solar Magnetohydrodynamics} (Kluwer Academic
Publishers,  Dordrecht, 2000).

\bibitem{Aschwanden2019-book-5}
M.~J. Aschwanden, \emph{New Millenium Solar Physics} (Springer,
Berlin, 2019).

\bibitem{Daughton2011POP}
H. Ji and W. Daughton,  Phys. Plasmas \textbf{18}, 111207 (2011).

\bibitem{Nakamura2016POP}
T. K. M. Nakamura, R. Nakamura, Y. Narita, W. Baumjohann, and W.
Daughton,  Phys. Plasmas \textbf{23}, 052116 (2016).

\bibitem{Chen2017POP}
J. Chen,  Phys. Plasmas \textbf{24}, 090501 (2017).

\bibitem{Zhelyazkov2015-6}
T.~V. Zaqarashvili, I.~Zhelyazkov, L.~Ofman,  Astrophys. J.
\textbf{813}, 123 (2015).

\bibitem{Solovev2014-7}
A.~Solov'ev, K.~Murawski, Astrophys. Space Sci. \textbf{350}, 11
(2014).

\bibitem{Liu2017-10}
Y.~Liu, X. Sun, T. T\"{o}r\"{o}k, V. S. Titov, and J E. Leake,
Astrophys. J. Lett. \textbf{846}, L6 (2017).

\bibitem{Titov2018-11}
V.~S. Titov,  C. Downs , Z. Miki\'{c}, T. T\"{o}r\"{o}k, J. A.
Linker, and R. M. Caplan, Astrophys. J. Lett. \textbf{852}, L21
(2018).

\bibitem{Cheremnykh2018-12}
O.~K. Cheremnykh, A.~N. Kryshtal, A.~A. Tkachenko, Adv. Space Res.
\textbf{61}, 603 (2018).

\bibitem{Zaitsev2020-13}
V.~V. Zaitsev, A.~V. Stepanov , P.~V. Kronshtadov, Solar Phys.
\textbf{295}, 166 (2020).

\bibitem{Parker2007-book-14}
E.~N. Parker, \emph{Conversations on Electric and Magnetic Fields
in the Cosmos} (Princeton University Press, 2007).

\bibitem{Melrose2017-15}
D.~B. Melrose, J. Geophys. Res.: Space Phys. \textbf{122}, 7963
(2017).

\bibitem{Solovev2021-16}
A.~A. Solov'ev, E.~A. Kirichek, Mon. Not. R. Astron. Soc.
\textbf{505}, 4406 (2021).

\bibitem{Kadomtsev1988-17}
B.~B. Kadomtsev, \emph{Tokamak Plasma: A Complex System} (IOP,
Bristol, 1992).

\bibitem{Miyamoto-18}
K. Miyamoto, \emph{Fundamentals of Plasma Physics and Controlled
Fusion} (Springer, Berlin, 2005)

\bibitem{Kadomtsev1987-19}
B.~B. Kadomtsev,  Sov. J. Plasma Phys.  \textbf{13}, 443 (1987).

\bibitem{Troyon1999-20}
F.~Troyon, R. Gruber, H. Saurenmann, S. Semenzato,  S. Succi,
Plasma Phys. Controlled Fusion \textbf{26}, 209 (1984).

\end{thebibliography}
\end{document}